\documentclass[conference]{IEEEtran}
\IEEEoverridecommandlockouts
\usepackage{cite}
\usepackage{amsmath,amssymb,amsfonts}
\usepackage{algorithmicx}
\usepackage{algorithm}
\usepackage{algpseudocode}
\usepackage{graphicx}
\usepackage{textcomp}
\usepackage{xcolor}
\usepackage{booktabs}
\usepackage{multirow}
\usepackage{marvosym}
\usepackage[colorlinks=true,urlcolor=black,citecolor=black,linkcolor=black]{hyperref}
\def\BibTeX{{\rm B\kern-.05em{\sc i\kern-.025em b}\kern-.08em
    T\kern-.1667em\lower.7ex\hbox{E}\kern-.125emX}}
\begin{document}

\title{Your Demands Deserve More Bits: Referring Semantic Image Compression at Ultra-low Bitrate
	\thanks{\Letter Corresponding author.}
	\thanks{This work was supported in part by the National Natural Science Foundation of China under Grant 62271119, the Natural Science Foundation of Sichuan Province under Grant 2025ZNSFSC0475 and the Independent Research Project of Civil Aviation Flight Technology and Flight Safety Key Laboratory under Grant FZ2022ZZ06.}
}

\author{
	\IEEEauthorblockN{Chenhao Wu$^*$, Qingbo Wu$^\dag$\textsuperscript{\Letter}, Haoran Wei$^*$, Shuai Chen$^*$, Mingzhou He$^*$, \\King Ngi Ngan$^\dag$, Fanman Meng$^\dag$, Hongliang Li$^\dag$}
	\IEEEauthorblockA{\textit{School of Information and Communication Engineering}\\
					\textit{University of Electronic Science and Technology of China}\\
				Chengdu, China\\
			$^*$\{chwu, hrwei, schen, hiram\}@std.uestc.edu.cn, $^\dag$\{qbwu, knngan, fmmeng, hlli\}@uestc.edu.cn}
}

\maketitle

\begin{abstract}
	With the help of powerful generative models, Semantic Image Compression (SIC) has achieved impressive performance at ultra-low bitrate. However, due to coarse-grained visual-semantic alignment and inherent randomness, the reliability of SIC is seriously concerned for reconstructing completely different object instances, even they are semantically consistent with original images. To tackle this issue, we propose a novel Referring Semantic Image Compression (RSIC) framework to improve the fidelity of user-specified content while retaining extreme compression ratios. Specifically, RSIC consists of three modules: Global Description Encoding (GDE), Referring Guidance Encoding (RGE), and Guided Generative Decoding (GGD). GDE and RGE encode global semantic information and local features, respectively, while GGD handles the non-uniformly guided generative process based on the encoded information. In this way, our RSIC achieves flexible customized compression according to user demands, which better balance the local fidelity, global realism, semantic alignment, and bit overhead. Extensive experiments on three datasets verify the compression efficiency and flexibility of the proposed method.
\end{abstract}


\section{Introduction}
Image compression is a fundamental topic in the multimedia field, aimed at transmitting and storing as much image information as possible while minimizing bitrate. Advanced image/video compression standards \cite{wallace1992jpeg,sullivan2012overview,bross2021overview} have significantly promoted the multimedia applications in the past two decades. However, they still struggle to reconstruct high-quality images at ultra-low bitrates. The emergence of Semantic Image Compression (SIC) algorithms \cite{bousselham2023gem,li2021cross,gao2023cross,lei2023text+sketch,careil2024towards} offers a promising solution to this challenge. For instance, PICS \cite{lei2023text+sketch} represents images as text descriptions and sketches, utilizing the pre-trained conditional diffusion model ControlNet \cite{zhang2023adding} for image decoding. PerCo \cite{careil2024towards} jointly trains an image compression codec with Stable Diffusion \cite{rombach2021highresolution} to encode images into text descriptions and limited image features. These approaches demonstrate superior realism and semantic consistency, even when the compression ratio exceeds 1000 times ($<0.024$bpp).

\begin{figure}
	\centering
	\includegraphics[width=0.8\linewidth]{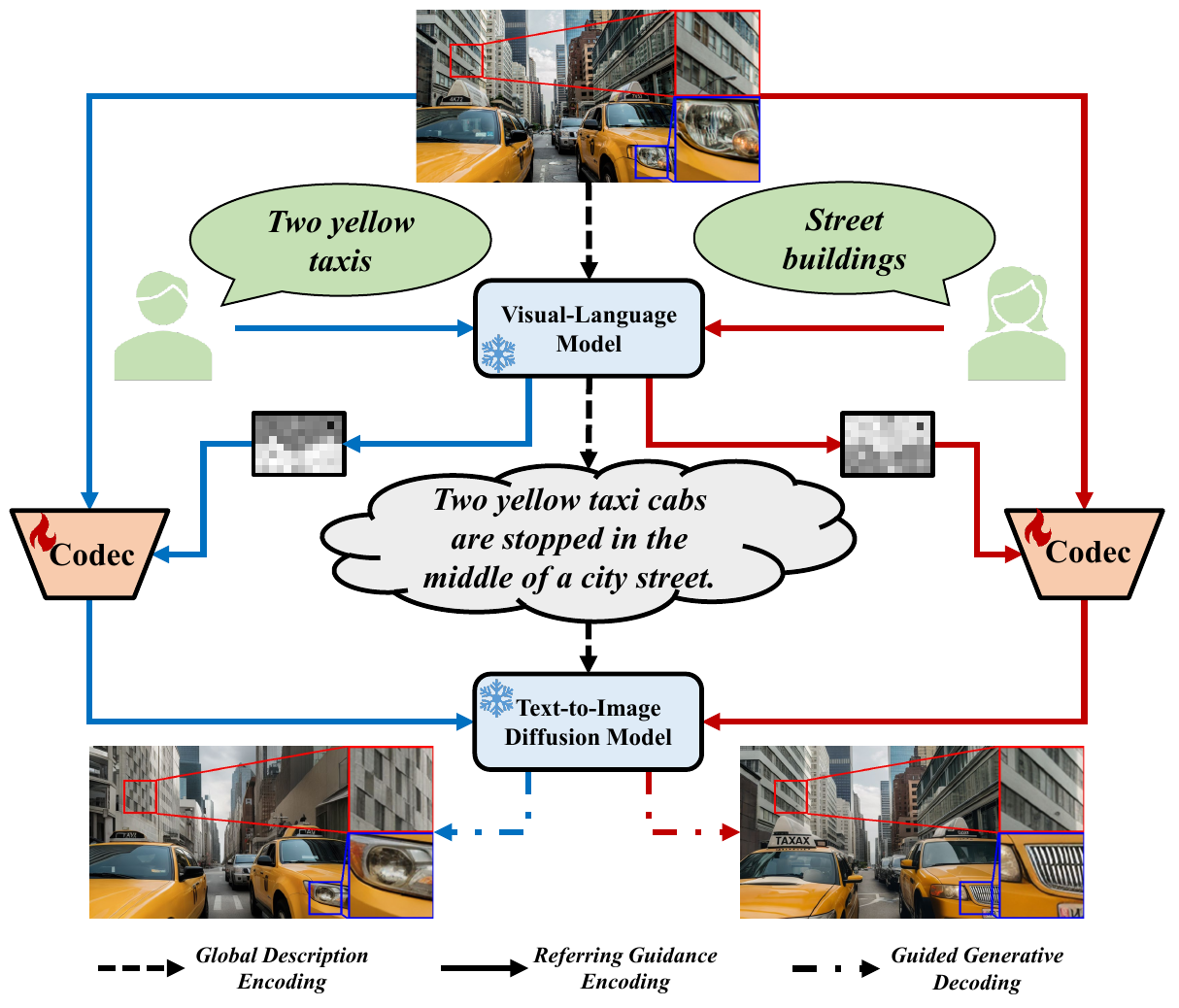}
	\caption{The proposed RSIC can improve the fidelity of user-specified content while performing ultra-low bitrate semantic compression.}
	\label{fig:head_framework}
\end{figure}

Despite their success in achieving extreme compression ratios, existing SIC methods easily decode the same semantic information into entirely different object instances due to coarse-grained visual-semantic alignment and inherent randomness. This limitation significantly restricts their applications in storing or transmitting important information as demanded by users.

To tackle this issue, we propose a novel Referring Semantic Image Compression (RSIC) framework. RSIC comprises three modules: Global Description Encoding (GDE) and Referring Guidance Encoding (RGE) in the encoder, along with Guided Generative Decoding (GGD) in the decoder. As shown in Fig. \ref{fig:head_framework}, GDE encodes the global description of the input image using a visual-language model to provide global semantic informtion. RGE encodes spatially non-uniform features based on user demands, offering guidance for local fidelity. GGD utilizes the global description as a condition and the local features as a guide, and employ a pre-trained diffusion model for image reconstruction. Throughout this process, the balance between global conditions and local guidance is adjusted according to user demands.

Our contributions can be summarized as follows:
\begin{itemize}
	\item We propose the novel referring semantic image compression framework that enhances the practicability of ultra-low bitrate image compression.
	\item Our method is plug-and-play compatible with off-the-shelf diffusion models, requiring no joint training or fine-tuning of pre-trained weights.
	\item Qualitative and quantitative experiments demonstrate the flexibility and superior performance of the proposed RSIC method across various user demands.
\end{itemize}

\section{Referring Semantic Image Compression}

\subsection{Global Description Encoding}
Global description can efficiently abstract and summarize image content into categories and attributes, and represent images at ultra-low bitrates. To accurately obtain the global description, we utilize the sophisticated visual-language model LLaVA \cite{liu2023llava,liu2023improvedllava}. Specifically, we input an image into LLaVA and prompt it with the query: \textit{Please describe this image to convey all semantic information}. After recieving the response, we use the text encoder zlib \cite{zlibHome74:online} to losslessly compress the global description, which typically requires about $60$ bytes.

\subsection{Referring Guidance Encoding}
In addition to the global description, local features can more effectively capture details that are challenging to describe in natural language, such as color, shape, and texture. Encoding additional features in specific local areas consumes more bitrate but offers improved fidelity. 

Based on user demands, we utilize the visual-language model GEM \cite{bousselham2023gem} to locate the user-specified content and generate a spatial weight map $M$. This map not only aids in bitrate allocation within the encoder but also balances fidelity and semantic alignment during decoding. To minimize the bitrate required for encoding $M$, we encode it at a spatial size of $1/64 \times 1/64$ of the original image, and quantize it to $1 \sim 8$ levels as needed. Without further compression, $M$ consumes a maximum of 0.0007 bpp, which remains acceptable even in ultra-low bitrate scenarios.

\begin{figure}
	\centering
	\includegraphics[width=0.8\linewidth]{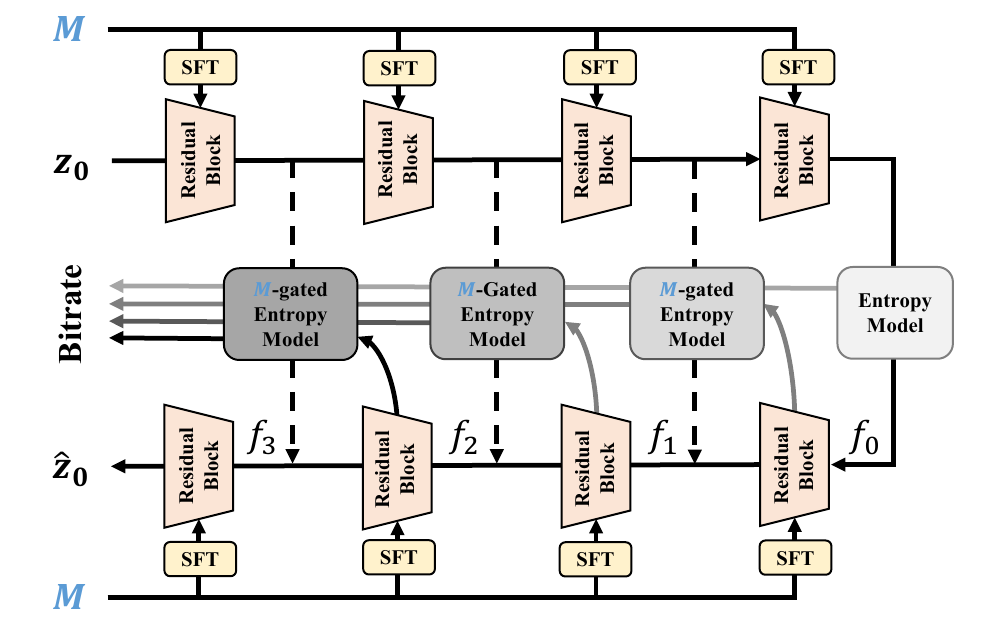}
	\caption{Structure of the proposed Hierarchical Spatially Variable-rate Latent Codec (HSVLC).}
	\label{compression_codec}
\end{figure}

Before applying compression, RSIC follows \cite{careil2024towards} and employs the encoder $\mathcal{E}$ of Stable Diffusion to reduce the image dimensions from $x \in \mathbb{R}^{H \times W \times 3}$ to $z_0 \in \mathbb{R}^{H/8 \times W/8 \times 4}$. Then we introduce a spatially variable-rate latent compression codec $\mathcal{C}$ to compress $z_0$ to $\hat{z}_0 = \mathcal{C}(z_0)$. Spatial Feature Transformation (SFT)\cite{song2021variable} has been proven effective in spatially variable-rate image compression, we further propose a Hierarchical Spatially Variable-rate Latent Coding (HSVLC) framework to improve its performance in latent space, as shown in Fig. \ref{compression_codec}.

In HSVLC, each residual block in the encoder/decoder contains a $2 \times$ downsampling/upsampling operation, resulting in four distinct coding scales. The $M$-gated entropy model controls the representation scale of each region. To further reduce bitrate consumption, we introduce the conditional prior between these scale representations following \cite{wu2025high}. Empirically, we set the boudaries between representation scales at $1/2$, $3/4$, and $7/8$, with $M$ ranging from $0$ to $1$.

To ensure ease of use and plug-and-play compatibility of the referring guidance, we train HSVLC directly on image features in latent space. Specifically, HSVLC takes $z_0$ as input, aiming to minimize both the distortion between the reconstruction $\hat{z}_0$ and $z_0$, as well as the total bitrate:
\begin{equation}
	\mathcal{L} = \sum_{i=0}^{i=3} \mathcal{R}(f_i) + \lambda \mathcal{D}(z_0, \hat{z}_0),
\end{equation}
where $\mathcal{R}(f_i)$ denotes the bit consumption of $i$-th representation scale, and $\mathcal{D}$ represents the distortion. We use Mean Square Error (MSE) as the distortion metric during training. 

\subsection{Guided Generative Decoding}

On the basis of the encoded global description and referring guidance, we use the pretrained text-to-image Stable Diffusion model \cite{rombach2021highresolution} for guided generative decoding. Specifically, the generative process of diffusion model is a reverse denoising process, which estimates and removes the additive noise $\tilde{\epsilon}_t$ at diffusion step $t$. Our global description condition can be directly injected through classifier-free guidance \cite{ho2021classifier}:
\begin{equation}
	\tilde{\epsilon}_t(\tilde{z}_t, c) = \epsilon_\theta(\tilde{z}_t, t) + \omega(\epsilon_\theta(\tilde{z}_t, c, t) - \epsilon_\theta(\tilde{z}_t, t)),
\end{equation}
where $c$ represents the global description, $\omega$ stands for the guidance scale, and $\epsilon_\theta$ is the noise predicted by U-Net.

\begin{figure}
	\centering
	\includegraphics[width=0.85\linewidth]{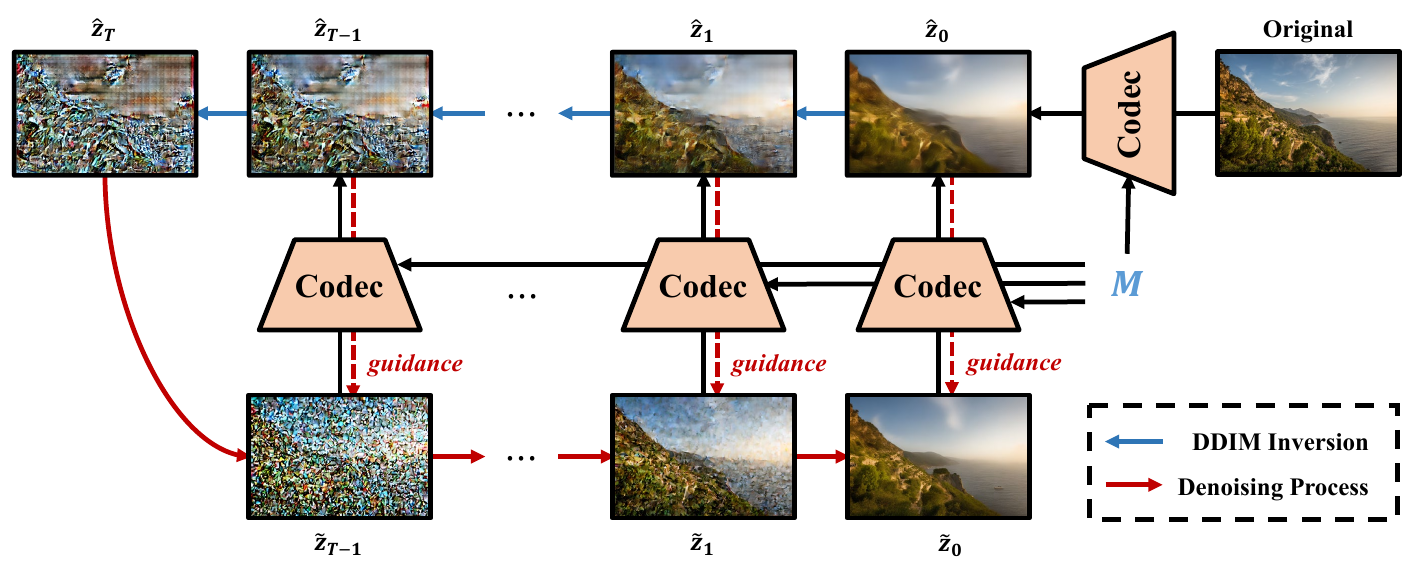}
	\caption{Guided generative decoding process.}
	\label{fig:compression_guidance}
\end{figure}

From the perspective of score-based generative modeling \cite{song2020score}, the unconditioned estimated additive noise $\epsilon_\theta(\tilde{z}_t,t) \approx \nabla \log p(\tilde{z}_t)$. To guide the generated results from random content to align with the referring guidance, $\nabla \log p(\tilde{z}_t)$ should be modified to $\nabla \log p(\tilde{z}_t|\hat{z}_0)$. Based on the assumption that the denosing process can be reversed in the limit of small steps \cite{mokady2023null}, $\hat{z}_t$ can be determined by DDIM Inversion \cite{mokady2023null} from $\hat{z}_0$. Following Bayes' theorem, we can deduce:
\begin{equation}
	\begin{aligned}
		\nabla \log p(\tilde{z}_t|\hat{z}_0) &= \nabla \log p(\tilde{z}_t|\hat{z}_t) \\
		&= \nabla \log p(\tilde{z}_t) + \nabla \log p(\hat{z}_t|\tilde{z}_t).
	\end{aligned}
\end{equation}
Therefore, the referring guidance can be injected by maximizing the probability of $\hat{z}_t$ given $\tilde{z}_t$. We achieve this by minimizing the distortion between $\hat{z}_t$ and $\mathcal{C}(\tilde{z}_t)$. The guidance process during the denoising phase is shown in Fig. \ref{fig:compression_guidance}. Please note that the reverse diffusion process is performed in latent space, while the features in Fig. \ref{fig:compression_guidance} are mapped into image space for better clarity.

\begin{algorithm}[b]\footnotesize
	\renewcommand{\algorithmicrequire}{\textbf{Input:}}
	\renewcommand{\algorithmicensure}{\textbf{Output:}}
	\caption{Guided Generative Decoding}
	\label{alg:compression_guidance}
	\begin{algorithmic}[1]
		\Require Image $x$, autoencoder $\mathcal{E}, \mathcal{D}$, image compression codec $\mathcal{C}$, diffusion steps $T$, self-recurrence steps $T_r$, referring scale $\gamma$, global description $c$.
		\State \textbf{Initialization:} Latent feature $\hat{z}_0=\mathcal{C}(\mathcal{E}(x))$, timestep $t=0$.
		\For{$t \in [0, T)$}
		\State $\hat{z}_{t+1 }= \sqrt{\bar{\alpha}_{t+1}/\bar{\alpha}_t}\hat{z}_t + \sqrt{\bar{\alpha}_{t+1}}(\sqrt{1/\bar{\alpha}_{t+1}-1}-\sqrt{1/\bar{\alpha}_t-1}) \cdot \tilde{\epsilon}_t(\hat{z}_t,c)$ \Comment{DDIM Inversion}
		\EndFor
		\State $\tilde{z}_T = \hat{z}_T$.
		\For{$t \in [T, 0)$}
		\For{$t_r \in [0, T_r)$}
		\State $\tilde{z}_{t-1} = \sqrt{\bar{\alpha}_{t-1}/\bar{\alpha}_t}\tilde{z}_{t} + \sqrt{\bar{\alpha}_{t-1}}(\sqrt{1/\bar{\alpha}_{t-1}-1}-\sqrt{1/\bar{\alpha}_t-1}) \cdot \tilde{\epsilon}_t(\tilde{z}_t,c) + \gamma \nabla \Vert \mathcal{C}(\tilde{z}_t) - \hat{z}_t \Vert_2$. \Comment{Denoising}
		\State $\tilde{z}_t$ = $\sqrt{\bar{\alpha}_t/\bar{\alpha}_{t-1}}\tilde{z}_{t-1} + \sqrt{1 - \bar{\alpha}_t/\bar{\alpha}_{t-1}} \cdot \epsilon'$ \Comment{Self-recurrence}
		\EndFor
		\EndFor
		\Ensure $\mathcal{D}(\tilde{z}_0)$
	\end{algorithmic}
\end{algorithm}

During generative decoding, RSIC adjusts the weights of two guidance to maintain a balance between fidelity and semantic consistency based on user demands. Specifically, RSIC employs a spatially non-uniform guidance scale defined as $\omega=3 \times (1 - M * 0.7)$. This approach ensures low global semantic guidance in regions with high $M$ values. Additionally, regions with larger $M$ values retain more information through the compression codec, and vice versa, thereby implicitly adjusting the referring guidance.

As observed in \cite{lugmayr2022repaint,bansal2023universal}, a sweet spot that ensures both realism and satisfaction of guidance constraints does not always exist for complex guidance functions. Therefore, we adopt the per-step self-recurrence method following \cite{bansal2023universal} by re-inject random gaussian noise $\epsilon' \sim \mathcal{N}(0, I)$ to $\tilde{z}_{t-1}$ to obtain $\tilde{z}_t$. The procedure of guided generative decoding is shown in Algorithm \ref{alg:compression_guidance}. 

\section{Experiments}

\begin{figure}
	\centering
	\includegraphics[width=\linewidth]{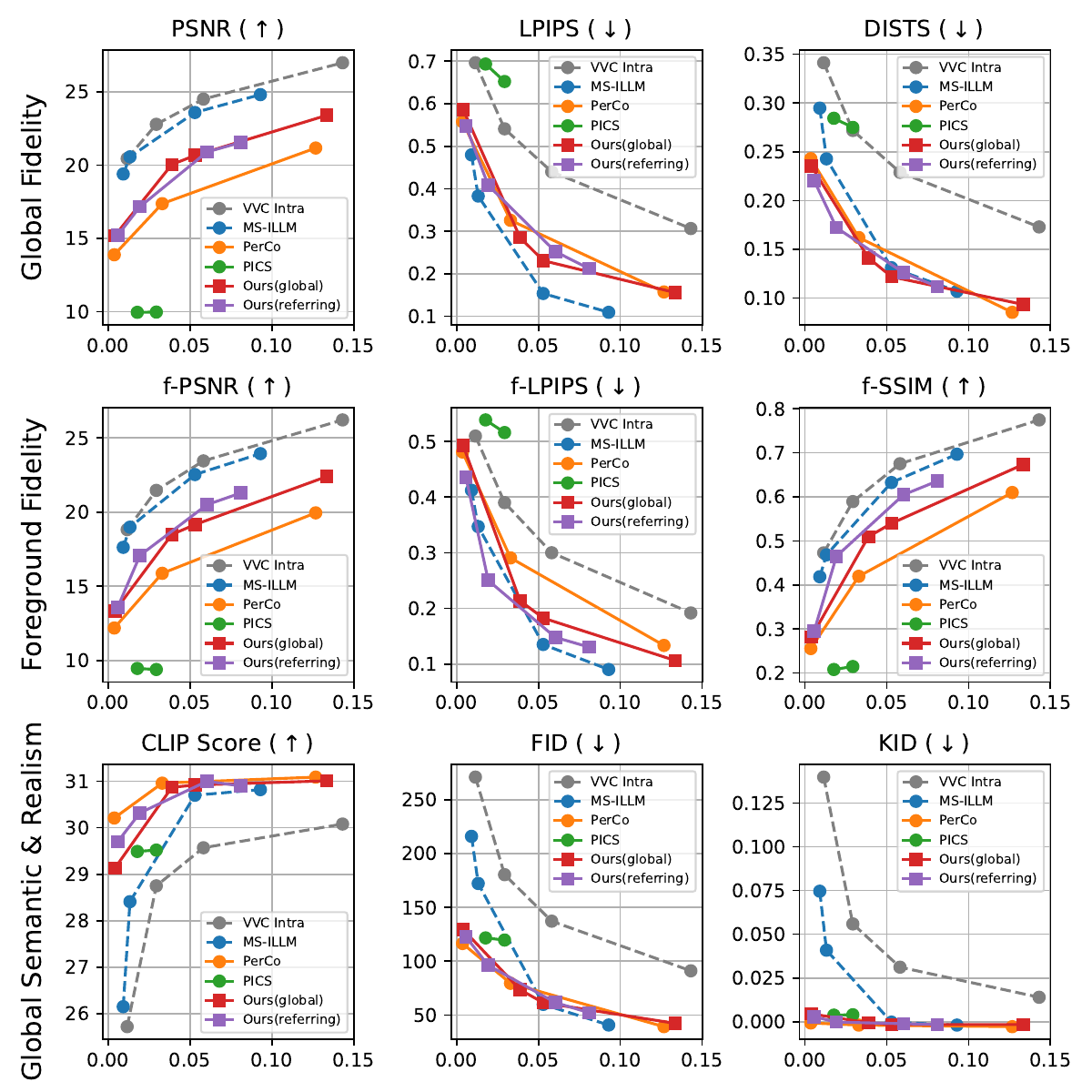}
	\caption{Evaluation of codecs on images from MS-COCO subset.}
	\label{fig:quantitative}
\end{figure}

\begin{figure*}
	\centering
	\includegraphics[width=0.95\linewidth]{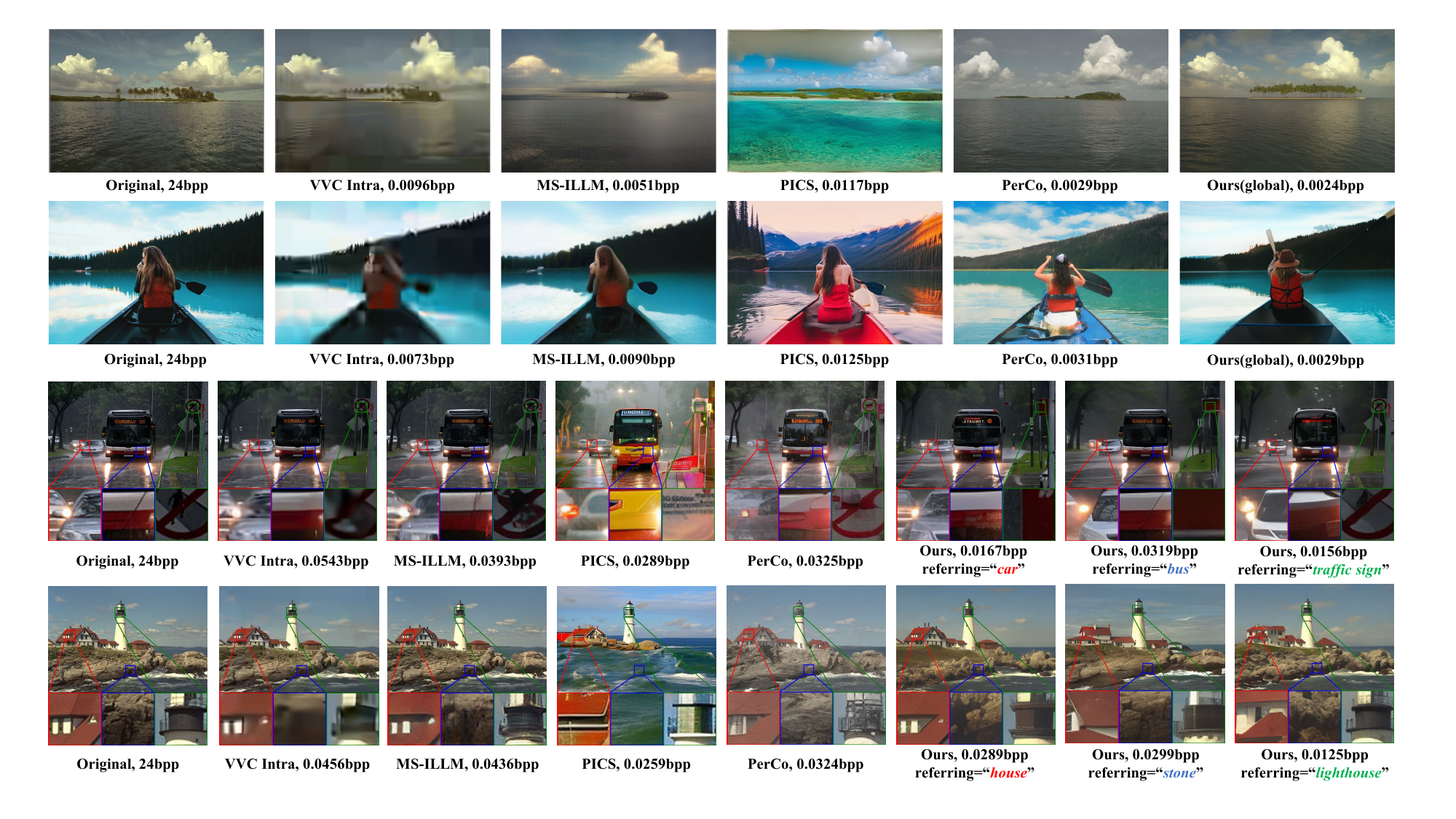}
	\caption{Qualitative comparison of the proposed method with other compression codecs, test images are selected from Kodak, CLIC, and MS-COCO datasets.}
	\label{fig:qualitative}
\end{figure*}

\subsection{Experimental Settings}
\textbf{Implementation details.} Our RSIC is implemented in both \textit{global} and \textit{referring} modes. In the \textit{global} mode, a spatially consistent weight map is employed, while in the \textit{referring} mode, the ground truth category label or user-provided description serves as the referring word to generate the spatial weight map. The training of HSVLC follows the settings in \cite{song2021variable}, with $\lambda=0.01\exp^{7M}$. We employ the deterministic DDIM sampling \cite{song2020denoising} to decode images, with diffusion sampling steps $T$ set to $50$. Self-recurrence steps $T_r$ is set to $8$, referring scale $\gamma$ is set to $10^3 \cdot \sqrt{\bar{\alpha}_{t-1}/\bar{\alpha}_t}$.

\textbf{Datasets.} Validation dataset of \textit{Open Images} \cite{kuznetsova2020open} is adopted as training dataset. The evaluation is performed on MS-COCO subset, Kodak dataset\cite{TrueColo62:online}, and CLIC dataset\cite{CLIC2020}. The MS-COCO subset consists of 226 images from MS-COCO validation dataset \cite{lin2014microsoft} with foreground proportions of $5\% \sim 25\%$. 

\textbf{Evaluation metrics.} We utilize FID \cite{heusel2017gans} and KID \cite{binkowski2018demystifying} to assess global realism, measuring the consistency of the distribution between the decoded image and the input image. The CLIP Score \cite{hessel2021clipscore} is employed to evaluate semantic alignment between decoded image and ground truth caption. For fidelity assessment, we use PSNR, LPIPS \cite{zhang2018unreasonable}, and DISTS \cite{ding2020iqa}. To evaluate the performance of referring coding, we calculate PSNR, LPIPS, and SSIM within foreground regions following \cite{guo2021image,guerreiro2023pct}, which are denoted as f-PSNR, f-LPIPS and f-SSIM.

\textbf{Baselines.} We conduct comparative experiments with traditional codec VVC Intra \cite{bross2021overview} and the learning-based codec MS-ILLM \cite{muckley2023improving}. Additionally, we compare our approach with diffusion-based SIC methods, PICS \cite{lei2023text+sketch} and PerCo \cite{careil2024towards}. Notably, neither the proposed method nor PICS alters the parameters of the diffusion model, while PerCo fine-tunes it.

\subsection{Quantitative Results}

Fig. \ref{fig:quantitative} presents the performance of the proposed RSIC compared to other approaches at ultra-low bitrates. In the figure, VVC Intra and MS-ILLM demonstrate superior fidelity but exhibit inferior realism and semantic alignment, while PerCo and PICS show the opposite trend. Our method performs well across all evaluation metrics. Specifically, in terms of realism and semantic alignment, RSIC matches the performance of the state-of-the-art algorithm PerCo in both global and referring modes, significantly surpassing VVC Intra and MS-ILLM at ultra-low bitrates. In terms of fidelity, our method notably enhances pixel fidelity (PSNR, f-PSNR) compared to PerCo and PICS, and establishes a leading position in perceptual fidelity (DISTS, f-LPIPS). Additionally, the comparison between the global mode and the referring mode indicates that RSIC effectively improves the fidelity of the focus area based on user demands.

\subsection{Qualitative Results}
We also present qualitative results in Fig. \ref{fig:qualitative}. The first two rows show the performance of our method compared to other approaches in \textit{global} mode. While VVC Intra and MS-ILLM maintain good consistency with the original image content, they exhibit noticeable unreal textures. In contrast, PICS and PerCo achieve commendable realism but lack fidelity to the original image. Our approach excels in both realism and fidelity with lower bitrates. The last two rows compare our method with other approaches in \textit{referring} mode. When users focus on specific elements, such as the \textit{traffic sign} or the \textit{lighthouse}, the fidelity of those elements is significantly enhanced.

\subsection{Ablations}

\begin{table}[]
	\centering
	\caption{Ablation study on referring semantic image compression.}
	\label{tab:ablation}
	\resizebox{0.8\linewidth}{!}{
		\begin{tabular}{c|cccc}
			\toprule
			Method & Bitrate & FID ($\downarrow$) & CLIP Score ($\uparrow$) & f-LPIPS ($\downarrow$) \\ \midrule
			W/o GDE & 0.0179 & 132.82 & 26.83 & 0.2665 \\ 
			W/o RGE	& 0.0018 & 133.16 & 28.19 & 0.6253 \\ 
			W/o HSVLC & 0.0187 & 100.06 & 30.18 & 0.3136 \\\midrule
			GGD: $T=10$ & 0.0197 & 149.67 & 28.95 & 0.4092 \\
			GGD: $T=30$ & 0.0197 & 101.02 & 30.31 & 0.2715 \\ \midrule
			Ours(referring) & 0.0197 & \textbf{96.22} & \textbf{30.32} & \textbf{0.2505} \\ \bottomrule
		\end{tabular}
	}
\end{table}

Table \ref{tab:ablation} presents the ablation results for each module in RSIC. Specifically, we analyse the effects of removing the global description (w/o GDE), omitting the referring guidance (w/o RGE), and changing the hierarchical codec to a single-layer entropy coding (w/o HSVLC). The results show that while the global description consumes limited bits, it significantly impacts realism and semantic alignment. Referring guidance acounts for the majority of the bitrate, directly affecting the fidelity of the reconstruction results. The hierarchical codec can enhance fidelity to a considerable extent. Additionally, we present the impact of sampling steps in the guided generative decoding process. 

\section{Conclusion}
This paper introduces a novel Referring Semantic Image Compression (RSIC) framework based on global description encoding, referring guidance encoding, and guided generative decoding. It is designed to balance local fidelity, global realism, semantic alignment, and bit consumption according to user demands. The effectiveness and flexibility of the proposed method are validated through extensive experiments. By injecting user demands into semantic image compression (SIC), the proposed framework effectively alleviates concerns about the unreliability of SIC, thereby enhancing the practicability of image compression at ultra-low bitrates.

\bibliographystyle{IEEEtran}
\bibliography{main_bib}

\end{document}